\documentstyle[preprint,aps,times]{revtex}

\newcommand{\bx}{{\bf x}}
\newcommand{\bu}{{\bf u}}
\newcommand{\ea}{{\bf e}_a}
\newcommand{\eq}{{\rm (eq)}}

\begin{document}
\draft
\title{Simulation of Rayleigh-B\'enard convection using lattice
Boltzmann method}
\author{Xiaowen Shan}
\address{USAF Phillips Laboratory, Hanscom AFB, Massachusetts 01731\\
Theoretical Division, Los Alamos National Laboratory, Los Alamos, New
Mexico 87545}
\date{\today}
\maketitle

\begin{abstract}
Rayleigh-B\'enard convection is numerically simulated in two- and
three-dimensions using a recently developed two-component lattice
Boltzmann equation (LBE) method.  The density field of the second
component, which evolves according to the advection-diffusion equation
of a passive-scalar, is used to simulate the temperature field.  A
body force proportional to the temperature is applied, and the system
satisfies the Boussinesq equation except for a slight compressibility.
A no-slip, isothermal boundary condition is imposed in the vertical
direction, and periodic boundary conditions are used in horizontal
directions.  The critical Rayleigh number for the onset of the
Rayleigh-B\'enard convection agrees with the theoretical prediction.
As the Rayleigh number is increased higher, the steady two-dimensional
convection rolls become unstable.  The wavy instability and aperiodic
motion observed, as well as the Nusselt number as a function of the
Rayleigh number, are in good agreement with experimental observations
and theoretical predictions.  The LBE model is found to be efficient,
accurate, and numerically stable for the simulation of fluid flows
with heat and mass transfer.
\end{abstract}

\section{Introduction}

Recently the lattice Boltzmann equation (LBE) method has been
developed as a new computational fluid dynamics (CFD) method.  This
method originated form a boolean fluid model known as the Lattice Gas
Automata (LGA) \cite{Frisch86,Wolfram86} which simulates the motion of
fluids by particles moving and colliding on a regular lattice.  The
averaged fluid variables, such as the density and velocity, were shown
to satisfy equations similar to the Navier-Stokes equations.  The LBE
method improves this idea by following only the ensemble-averaged
distribution functions, therefore eliminating the time-consuming
statistical average step in the original LGA \cite{McNamara88}.
Simplified collision models were later used in place of the collision
operator derived from the LGA to improve both the computational
efficiency and the accuracy.  Most noteworthy, the simple collision
model of Bhatnagar, Gross, and Krook \cite{Bhatnagar54} was applied to
the lattice Boltzmann equation, yielding the so called lattice BGK
model \cite{Chen92a,Qian92a}.  The additional flexibility in this
approach allows the removal of the artifacts of the LGA, specifically
the lack of Galilean invariance and the velocity dependent pressure.
This method was found numerically to be at least as stable, accurate
and computationally efficient as traditional CFD methods for
simulation of simple single-phase incompressible flows
\cite{Chen92b,Martinez94a,Hou95}.  More importantly, since fluid
motion is simulated at the level of the distribution functions, the
microscopic physics of the fluid particles can be incorporate easily
like in other particle methods.  Many complex fluid phenomena due to
interparticle interactions, such as capillary phenomena, multiple
phase flows, and non-linear diffusion, can be simulated naturally
\cite{Shan93,Shan94,Martys96}.

In most LBE models so far, only mass and momentum conservation is
implemented.  The macroscopic equations of these models correspond to
the Navier-Stokes equation with an ideal-gas equation of state and a
constant temperature.  However it is important and sometimes critical
to have the capability of simulating thermal effects simultaneously
with the fluid flows.  Obviously the temperature distribution in a
flow field is of central interest in heat transfer problems.  In most
geophysical flows, the temperature difference is the driving mechanism
of the motion of the fluid.  More importantly, when part of the fluid
system undergoes phase transition, as in the boiling and evaporation
processes, the evolution of the temperature field is directly coupled
with the fluid dynamics.  Since the LBE method has the most advantage
in the simulation of complex fluids with multiple phases and phase
transitions, it is necessary to develop the capability of simulating
thermodynamics with the LBE method.

In general, the simulation of thermal systems by the LBE method has
not achieved the same success as that of iso-thermal flows.
Theoretically, a LBE model with energy conservation can be constructed
\cite{Alexander93a,Chen95} to yield a temperature evolution equation
at the macroscopic level.  However, the model so obtained suffers from
severe numerical instability \cite{McNamara95}, especially in
three-dimensions (3D).  Additional stabilization procedures have to be
invoked to achieve stability comparable to that of conventional CFD
methods, {\em e.g.}, finite difference schemes.  Moreover, when
interparticle forces are included as in the multi-phase models, the
energy conservation is further complicated by the potential part of
the internal energy.  For these reasons constructing a practically
usable non-ideal-gas LBE model with energy conservation is difficult
if not impossible.  Nevertheless, in many circumstances where the
viscous and compressive heating effects can be neglected (small
Brinkman number limit), the temperature field is passively advected by
the fluid flow and obeys a much simpler passive-scalar equation.  This
same equation also governs the diffusion of each individual component
in a fluid mixture.  By taking advantage of this formal analogy
between heat and mass transfer, we can simulate the temperature field
as an additional component of the fluid system.  Early two-component
LGA model \cite{Burges87} exhibited qualitative features of thermal
convections.  In a previously developed multiple component LBE model
\cite{Shan95}, we have shown that the evolution of the concentration
fields is Galilean invariant and obeys Fick's law.  The diffusivity is
independent of the viscosity, allowing a changeable Schmidt number (or
Prandtl number in the terminology of heat transfer).  This model does
not implement energy conservation and therefore has the same stability
as the non-thermal LBE models and other conventional CFD methods.  By
adding one more component, the computation efficiency, either
memory-wise or time-wise, is not compromised compared with the
approach of direct implementation of energy conservation because fewer
speeds are required for each component.

In this paper, we present the simulation of the Rayleigh-B\'enard
convection (RBC) as an example.  Due to its simplicity and the
richness of the phenomena, this problem has been extensively studied
both theoretically and experimentally
\cite{Busse86,Domaradzki88,Massaioli93,Massaioli95,Siggia94} and
serves as an excellent benchmark problem for numerical schemes because
detailed results are available for comparison with numerical
computations.  In section~\ref{sec:2}, we briefly review the multiple
component LBE model and then formulated it for the simulation of the
Boussinesq equation.  The implementation of the isothermal no-slip
boundary condition is also discussed.  In section~\ref{sec:3},
simulation results are presented and compared with theoretical and
experimental results.  The limitation and some further extensions of
this method are discussed in section~\ref{sec:4}.

\section{Numerical method}
\label{sec:2}

The following single component lattice Boltzmann equation with BGK
collision term describes the evolution of the distribution function
$n_a(\bx, t)$ in space $\bx$ and time $t$.
\begin{equation}
n_a(\bx + \ea, t + 1) - n_a(\bx, t) = -\frac 1\tau\left[n_a(\bx, t)
- n_a^\eq(\bx, t)\right], \quad a = 1, \cdots, b\label{eq:lbe}
\end{equation}
The set of $b$ vectors $\{\ea; a = 1, \cdots, b\}$ pointing from each
lattice site to its neighboring sites forms the discretized velocity
space of the distribution function.  The macroscopic number density,
$n(\bx, t)$, and velocity, $\bu(\bx, t)$, of the fluid are obtained
from $n_a$ as $n = \sum_an_a$ and $n\bu = \sum_an_a\ea$.
Eq.~(\ref{eq:lbe}) represents the relaxation of the distribution
function to its equilibrium value, $n_a^\eq$, which is a function of
$n$ and $\bu$ only.  The choice of $n_a^\eq$ has to ensure that the
macroscopic fluid equation obtained from Eq.~(\ref{eq:lbe}) by
Chapman-Enskog calculation \cite{Chapman70} agrees with the
Navier-Stokes equations.  The functional form of $n_a^\eq$ depends on
the structure of the lattice and is usually not uniquely determined.
For square and cubic lattices in 2D and 3D, the following form of
$n_a^\eq$ was shown to yield Navier-Stokes equations by Qian {\em et
al.} \cite{Qian92a}:
\begin{equation}
n_a^\eq = w_an\left[1 + 3\ea\cdot\bu + \frac{9}{2}(\ea\cdot\bu)^2 -
\frac{3\bu\cdot\bu}{2}\right].\label{eq:feq}
\end{equation}
Here $w_a$ is a function of $|\ea|$ and depends on the number of
speeds included in the model.  In the present work, 9 and 15
velocities are used in 2D and 3D computation respectively.  The
$w_a$'s were given as 4/9, 1/9, and 1/36 for $|\ea|= 0, 1, \sqrt{2}$
in 2D and 2/9, 1/9, and 1/72 for $|\ea|= 0, 1, \sqrt{3}$ in 3D
\cite{Qian92a}.  It can be easily verify that the 2D distribution
function is a degenerate case of the 3D version if the flow is
two-dimensional.

\subsection{Multiple component LBE model}

The multiple component LBE model with interparticle interaction
\cite{Shan93} was originally developed for simulation of multi-phase
flows and phase transitions.  The components can be miscible or
partially immiscible depending on the strength of the interaction.
When the interaction is weak, or in a single phase region of a
multiphase system, this model can be used to simulate diffusion due
to various driving mechanisms \cite{Shan96}.  In this model, the
distribution function of each component evolves according to
Eq.~(\ref{eq:lbe}).  The same form of the equilibrium distribution
function given by Eq.~(\ref{eq:feq}) is used for all the components
except that $n$ and $\bu$ are calculated separately for each
component.  In the absence of any interaction and external forces, the
distribution functions of all the components were assumed to have a
common velocity, $\bu'$.  The conservation of the total momentum at
each collision requires that
\begin{equation}
\bu' = \sum_{\sigma=1}^S\frac{m_\sigma n_\sigma\bu_\sigma}
{\tau_\sigma}/\sum_{\sigma=1}^S
\frac{m_\sigma n_\sigma}{\tau_\sigma},\label{eq:uprime}
\end{equation}
where $S$ is the number of components in the system; $m_\sigma$,
$\tau_\sigma$ and $n_\sigma = \sum_an_a^\sigma$ are the molecular
mass, the relaxation time, the number density of the component
$\sigma$ respectively, and $m_\sigma n_\sigma\bu_\sigma =
m_\sigma\sum_an_a^\sigma\ea$ is the momentum of component~$\sigma$
calculated from its distribution function $n_a^\sigma$.  When the
force ${\bf F}_\sigma$ is applied to component $\sigma$, the momentum
has to be incremented correspondingly.  This was done by replacing
$\bu$ in Eq.~(\ref{eq:feq}) with $\bu' + \tau_\sigma{\bf
F}_\sigma/\rho_\sigma$.  The force ${\bf F}_\sigma$ in general
includes both interparticle forces and external forces.  For
nearest-neighbor interaction, the following form of the interparticle
force was proposed as it conserves the total momentum of the system
and yields an adjustable equation of state at the macroscopic level:
\begin{equation}
{\bf F}_\sigma = -\psi_\sigma\sum_{\bar{\sigma}}{\cal G}_{\sigma
\bar{\sigma}}\sum_a\psi_{\bar{\sigma}}(\bx+\ea)\ea,
\end{equation}
where $\psi_\sigma$ is an arbitrary function of the number density of
the $\sigma$th component.

In the most general multiple component LBE model with interparticle
interaction and external forces, there are three types of diffusions
due to different driving mechanisms \cite{Shan96}.  They are {\em
ordinary diffusion}, {\em pressure diffusion} and {\em forced
diffusion}.  With the equilibrium distribution functions given by
Eq.~(\ref{eq:feq}), the pressure diffusion does not appear; if a
common acceleration is applied to all the components, namely ${\bf
F}_\sigma = \rho_\sigma{\bf g}$, forced diffusion is also absent.  The
only type of diffusion left is the ordinary diffusion due to
concentration gradients which obeys Fick's law.  In addition, a
components, {\em e.g.} component~$S$, can be made to behave as a
passive-scalar by setting its molecular mass to zero together with its
interaction with all the other components, namely $m_S\rightarrow 0$
and ${\cal G}_{\sigma S}\rightarrow 0$ for $\sigma = 1, \cdots, S-1$.
This component will not contribute to the total momentum of the
mixture.  It is simply advected ``passively'' and diffuses into the
main flow, having no effect on the flow.

For the study of the RBC, we employ a two-component system;
component~1 represents the motion of the fluid and component~2
simulates a passive temperature field.  The distribution functions of
the two components evolve according to Eqs.~(\ref{eq:lbe}) and
(\ref{eq:feq}), with $\bu$ in Eq.~(\ref{eq:feq}) being replaced by
$\bu_1 + \tau_\sigma{\bf g}$ for both components.  Since the molecular
masses of the two components no longer appear in the dynamic
equations, they are set to unity.  The density and the fluid velocity
are calculated from the distribution function of component~1 as $\rho
= \sum_an_a^1$ and $\bu = \bu_1 + {\bf g}/2$,
(c.f. Ref.~\cite{Shan95}).  They satisfy the following equations
\begin{eqnarray}
&&\frac{\partial\rho}{\partial t} + \nabla\cdot(\rho\bu) = 0
\label{eq:ct}\\
&&\frac{\partial\bu}{\partial t} + \bu\cdot\nabla\bu =
-\frac{\nabla p}{\rho} + \nu\nabla^2\bu + {\bf g}.
\label{eq:ns}
\end{eqnarray}
where the pressure $p$ is related to $\rho$ by the equation of state
$p = \frac 13\rho + \frac 32{\cal G}_{11} \psi^2(\rho)$.  In the
simulation of RBC, it is sufficient to set ${\cal G}_{11} = 0$.  The
kinematic viscosity $\nu$ is given by
\begin{equation}
\nu = \frac 13\left(\tau_1 - \frac 12\right)
\label{eq:nu}
\end{equation}
as in the ordinary LBE models.  The number density of the second
component satisfy the following diffusion equation \cite{Shan96}:
\begin{equation}
\frac{\partial n_2}{\partial t} + \nabla\cdot(n_2\bu) =
\nabla\cdot({\cal D}\nabla n_2).
\end{equation}
The temperature field $\theta$ can be simulated by the density field
$n_2$.  When the compressibility is negligible as in the small Mach
number limit, the velocity field is approximately divergence-free and
the temperature field satisfies the following ``passive-scalar''
equation:
\begin{equation}
\frac{\partial\theta}{\partial t} + \bu\cdot\nabla\theta =
\nabla\cdot({\cal D}\nabla\theta)
\label{eq:a},
\end{equation}
where the diffusivity, ${\cal D}$, is given by
\begin{equation}
{\cal D} = \frac 13\left[\tau_2(1+9{\cal G}_{22}\psi_2
d\psi_2/dn_2) - \frac 12\right].\label{eq:D}
\end{equation}
The diffusivity can be tuned independently of the viscosity by
changing either $\tau_2$ or the interaction strength, ${\cal G}_{22}$.
For simplicity, ${\cal G}_{22}$ is also set to zero in the present
simulation.  The LBE model is a much simplified version of that in
\cite{Shan93} since no interparticle interaction is used.

\subsection{Simulation of the Rayleigh-B\'enard convection}

In the most common form of RBC, a thin layer of viscous fluid is
confined between two horizontal rigid boundaries maintained at
different temperatures.  When the fluid has a positive thermal
expansion coefficient, and the gravity is in the same direction of the
temperature gradient, the net buoyancy force is in the opposite
direction of the gravity.  As the temperature difference between the
two boundaries is raised above a certain threshold, the static
conductive state becomes unstable, and convection occurs abruptly.

The well-known Boussinesq approximation is often used in the study of
natural convection.  With this approximation, the material properties
are assumed to be independent of temperature except in the body force
term, where the fluid density $\rho$ is assumed to be a linear
function of the temperature, namely $\rho/\rho_\infty = 1 +
\beta(T-T_\infty)$.  Here $\rho_\infty$ and $T_\infty$ are the density
and temperature at the reference point, and $\beta$ the constant
thermal expansion coefficient.  The gravitational force is therefore
$\rho_\infty{\bf g} + \rho_\infty{\bf g}\beta(T - T_\infty)$.  After
absorbing the first term into the pressure, the effective body force
is linearly proportional to the temperature variation.

The Boussinesq equation can be simulated with the two-component LBE
model by making the external gravitational acceleration, ${\bf g}$, in
Eq.~(\ref{eq:ns}) a linear function of the temperature $\theta$, {\em
i.e.}, ${\bf g} = -g\theta{\bf e}_z$, where ${\bf e}_z$ is the unit
vector in the vertical direction and $g$ a parameter controlling the
strength of the force.  In Cartesian coordinates $(x, y, z)$, the
lattices are of the sizes $L_x\times L_z$ and $L_x\times L_y\times
L_z$ in 2D and 3D respectively.  Periodic boundary condition is used
in $x$ and $y$ directions, and the following no-slip, isothermal
boundary condition is used in $z$ direction:
\begin{equation}
\left\{\begin{array}{ll}
\bu = 0, \theta = 0 & z = 0;\\
\bu = 0, \theta = 1\quad & z = L_z.
\end{array}\right.\label{eq:bc}
\end{equation}

Since the LBE fluid is always compressible, an externally applied
force will cause a density variation.  This compressible effect can be
eliminated by absorbing into the pressure term the part of the body
force that corresponding to the body force in the static conductive
state, yielding the following net external acceleration
\begin{equation}
{\bf g} = -g\left(\theta - \frac{z}{L_z}\right){\bf e}_z.
\end{equation}
In the conductive state the above external force vanishes and the
density field is homogeneous.

For a given the system size, the characteristic velocity, the Grashof
number, the Rayleigh number and the Prandtl number are determined by
the three parameters $\tau_1$, $\tau_2$ and $g$ in the LBE model as
the following:
\begin{equation}
v_c = \sqrt{gL_z}, \quad Gr = \frac{gL_z^3}{\nu^2}, \quad R = GrPr =
\frac{gL_z^3}{\nu{\cal D}},\quad Pr = \frac{\nu}{\cal D} =
\frac{2\tau_1 - 1}{2\tau_2 - 1}.
\label{eq:param}
\end{equation}
The Prandtl number is determined by the two relaxation times used for
the two components.  Given the two basic characteristic dimensionless
numbers $Pr$ and $R$, there is an extra degree of freedom in
determining $\tau_1$, $\tau_2$ and $g$.  However, to ensure that the
Mach number is small, $v_c$ has to be kept small.  Once $v_c$ is
chosen, all the parameters in the LBE model are determined by the two
dimensionless numbers $R$ and $Pr$.

\subsection{Implementation of the boundary conditions}

To implement the isothermal, no-slip boundary condition, we must
ensure that at the boundary, the component simulating the fluid flow
has zero velocity, and the component simulating the temperature field
has fixed density.  The mass flux of the second component represents
the heat transport through the boundaries.  Usually the LGA and LBE
methods implement the no-slip boundary condition by reversing the
direction of the incoming particles at the boundary, yielding zero
averaged velocity.  This simple ``bounce-back'' method was found to be
inaccurate \cite{Cornubert91,Ginzbourg94}.  In the present work, it
results in errors of up to 50\% in the critical Rayleigh number.  More
accurate and general methods have been developed to implement velocity
boundary conditions in complex geometry
\cite{Noble95,Zou95,Ginzbourg96}.  These methods usually involve
additional computation at the boundary sites.  Here, because the
boundaries are flat planes, both the velocity and the density boundary
conditions can be implemented more efficiently.

When analyzing various implementations of boundary conditions, exact
solutions in some simple cases are found to be very useful
\cite{Zou95,He96}.  For simplicity, we consider the time-independent
one-dimensional situation.  All variables depend only on $z$, the
coordinate perpendicular to the wall, so that the spatial dependence
can be noted by a single superscript, $j$, starting from 0 at the
lower boundary.  The elements of the distribution functions $n^j_a$
can be classified into three groups, $n^j_+$, $n^j_-$ and $n^j_0$,
according to the sign of $\ea\cdot{\bf e}_z$.  Eq.~(\ref{eq:lbe})
reduces to the following simple form:
\begin{equation}
n_\pm^{j\pm 1} - n_\pm^j = -\frac{1}{\tau}\left[n_\pm^j -
n_\pm^{j\eq}\right] \quad {\rm and} \quad n^j_0 = n^{j\eq}_0.
\label{eq:bc1}
\end{equation}

We assume the distribution functions at all sites including the
boundary sites are updated uniformly using Eqs.~(\ref{eq:bc1}).  At
the lower boundary sites, the groups $n_+^0$ and $n_0^0$ are
unspecified.  The only available information about the bulk of the
fluid is $n^0_-$, from which, $n^1_+$ is to be constructed according
to some updating scheme so that certain hydrodynamic boundary
condition is satisfied at macroscopic level.  The ``bounce-back''
scheme simply sets $n^1_a = n^0_b$ for any $a$ and $b$ satisfying
${\bf e}_a = -{\bf e}_b$.  Obviously this in general does not satisfy
Eqs.~(\ref{eq:bc1}) with $\bu = 0$ at the boundary.  However, if we
use the ``bounce-back'' scheme to calculate the group $n_+^0$, namely
we set $n^0_a = n^0_b$ for any $a$ and $b$ satisfying ${\bf e}_a =
-{\bf e}_b$, and calculating $n_+^1$ using Eqs.~(\ref{eq:bc1}) with
$\bu = 0$ and $n = 6\sum n_-^0$ in the computation of $n^{0\eq}_+$,
the no-slip boundary condition will be satisfied.  Here the summation
is over all the elements in the group.

The isothermal boundary condition is imposed by fixing the density of
the second component at specified values on the boundaries.  In the
time-independent one-dimensional situation, the density profile of the
passively convected component can be exactly solved from
Eqs.~(\ref{eq:bc1}).  We sum all the elements of the distribution in
each of the groups and note the sum as $N^j_\pm = \sum n^j_\pm$.  By
summing Eqs.~(\ref{eq:bc1}) we find
\begin{equation}
N_\pm^{j\pm 1} - N_\pm^j = -\frac{1}{\tau}\left[N_\pm^j -
N_\pm^{j\eq}\right]
\end{equation}
Using Eq.~(\ref{eq:feq}) and notice that the velocity only has
component parallel to the wall, we can find easily that $N^{j\eq}_\pm
= n^j/6$, independent of the local velocity.  Here $n^j$ is the
density at the $j$-th position.  From the second part of
Eq.~(\ref{eq:bc1}), we have $N_+^j + N_-^j = n^j/3$.  For any three
consecutive $j$ values, there are total of seven equations relating
the six variables $N_\pm^j$ to the three density values.  On
eliminating $N_\pm^j$ from the seven equations, we find $2n^j = n^{j -
1} + n^{j + 1}$; namely, the density profile is linear in $z$ as the
diffusion equation predicts.

At the boundary sites, we must have
\begin{equation}
N_+^0 = N^0/3 - N^0_-.
\label{eq:bc2}
\end{equation}
In the computation, the isothermal boundary condition is implemented by
computing the distribution function elements in the group $n_+^0$
according to the following equation:
\begin{equation}
n_a^0 = 2w_an - n_b^0,
\end{equation}
and then updating it using Eq.~(\ref{eq:lbe}).  Here $a$ and $b$ are
any pair of indices such that ${\bf e}_a$ and ${\bf e}_b$ are mirror
images of each other with respect to the boundary.  In addition to
satisfying Eq.~(\ref{eq:bc2}), this scheme is also compatible with the
no-slip boundary condition.

\section{Simulation results}
\label{sec:3}

We present the simulation results in this section.  The two basic
characteristic dimensionless numbers are the Rayleigh number $R$ and
the Prandtl number $Pr$.  The velocity and time reported hereafter are
in the units of $v_c$ and the characteristic time, $L_z/v_c$,
respectively.  The diffusion time $t_d = L_z^2/{\cal D}$ is
$\sqrt{PrR}$ in these units.

\subsection{Onset of Rayleigh-B\'enard instability}

The critical Rayleigh number at which the static conductive state
becomes unstable was given by the linear stability theory and
confirmed by laboratory observations.  The static conductive state is
found to first become unstable to the disturbance of the wave number
$k_c$ = 3.117 in the $x$-$y$ plane when the Rayleigh number exceeds
the critical value of $R_c$ = 1707.762.  If the deviation from the
Boussinesq approximation is small, the convection occurs in the form
of two-dimensional rolls.  Since the development of the instability is
very slow at near-critical Rayleigh numbers, the computation has to be
carried out for a long time before stable convection is fully
developed.  Because the first unstable disturbance is two-dimensional,
we conduct the near-critical simulations primarily in 2D to save CPU
time.  The results were compared with 3D simulation results for some
typical cases.

With periodic boundary condition, the wave number in $x$-$y$ plane can
only take discretized values given by
\begin{equation}
\left(\frac{k}{2\pi L_z}\right)^2 = \left(\frac{n_x}{L_x}\right)^2
+ \left(\frac{n_y}{L_y}\right)^2, \quad n_x, n_y = 0, 1, 2, \cdots.
\end{equation}
In 2D, the aspect ratio $L_x/L_z$ has to be a multiplication of
$2\pi/k_c$ to accommodate the disturbance of the wave number $k_c$.
Of course this can only be satisfied approximately on a uniform
lattice.  In the near-critical computation, unless otherwise
specified, we chose $L_x \simeq 2\pi L_z/k_c$ to save computation
cost.

To measure the critical Rayleigh number, computations were started
from the static conductive state at several different Rayleigh numbers
close to $R_c$.  An initial small perturbation was applied to the
density field.  The growth rates of the disturbance were then measured
and extrapolated to obtain the Rayleigh number corresponding to zero
growth rate.

Shown in Fig.~\ref{fg:grow} are the typical time-histories of the
maximum velocities in $z$ direction for three slightly-above-critical
Rayleigh numbers of 1720, 1735 and 1750 respectively.  The other
parameters are $Pr$ = 1, $\tau_1$ = 1, and $L_z$ = 50 for all three
runs.  The peak velocity is found to grow exponentially and then
saturate at a finite amplitude.  The steady-state isotherms and the
velocity field are shown in Fig.~\ref{fg:tnv} for the simulation with
$R$ = 1750.  The growth rates were measured with least-square fitting
in the exponential growth stage.  The fitting results are shown as the
straight lines.

The measured growth rates were plotted against the Rayleigh number in
Fig.~\ref{fg:rates}.  Three sets of simulations with $\tau_1$ = 0.55,
1., and 1.5 were performed to investigate the accuracy of using
different values of $\tau_1$.  All other parameters were the same in
these simulations.  A solid straight line is fitted through the data
points for each set of data.  The intersections of the lines with the
$x$-axis give the Rayleigh numbers corresponding to neutral stability.
It is to be seen that near the critical Rayleigh number, using a
$\tau_1$ other than unity tends to change the growth rates, which
causes an error in the prediction of the critical Rayleigh number.

We have also measured the critical Rayleigh number for different
Prandtl numbers and with different system sizes.  The measurement
results and the parameters used are summarized in Table~\ref{tbl:rc}.
The biggest error seems to have been caused by using a large $\tau$
value in the computation.  Fortunately, this does not impose a
significant limitation on the range of physical parameters that can be
simulated, because for a given Rayleigh number, $\tau_1$ and $\tau_2$
can always be kept small by using a small $g$.

Shown in the first part of Table~\ref{tbl:rc} are five otherwise
identical runs with different system sizes.  The time history of the
peak vertical velocity in these runs are plotted in
Fig.~\ref{fg:conv}.  In the plot, the starting times were adjusted so
that the initial perturbation levels are the same for all five runs.
It can be seen that the convergence is fast and differences between
the runs with $L_z >$ 20 is insignificant.

Also shown in Fig.~\ref{fg:conv} is the result of a 3D simulation on a
$128\times 128\times 32$ lattice with the same parameters.  The growth
rate in the early stage is the same as that in the 2D simulations.
However, the peak velocity over-shoots before it saturates at the same
level.  Figs.~\ref{fg:temp} display a series of snapshots of the
temperature distribution on a $x$-$y$ plane laying in the middle
between the two walls at the times $t$ = 1047, 1320, 1524 and 2273.
The grey scales from the darkest to the brightest represents the
temperature in the range 0.403 $< \theta <$ 0.597.  The instability
starts in the form of an array of convection cells, the superposition
of the most unstable mode ($k = k_c$) oriented in the $x$ and $y$
directions, and reaches its maximum near $t$ = 1320
(Fig.~\ref{fg:temp}b).  The fully developed convection rolls oriented
in one direction seem to suppress the orthogonal rolls and the final
convection pattern is purely two-dimensional (Fig.~\ref{fg:temp}d).

\subsection{Higher Rayleigh number}

The two-dimensional convection pattern characterized by the rolls is
unstable at higher Rayleigh number.  As the Rayleigh number is being
increased, a series of transitions to more complicated states occur,
and the form of the convection becomes both three dimensional and
time-dependent, and eventually turbulent at very high Rayleigh number.
Detailed numerical simulation of all the complicated transitions and
the different forms of convection requires a large amount of
computation.  This is because the form of the convection depends on
both the initial condition and the boundary conditions.  A large number
of runs have to be performed to cover the parameter space.  In addition,
the computation has to be carried out for a long time due to the large
differences among the time scales in the problem.  Here we only present
the simulation results for a few typical situations at moderate
Rayleigh numbers due to the limitation of computer resources.

A two-dimensional simulation at high Rayleigh numbers was performed on
a 101 $\times$ 50 lattice with a Prandtl number of 0.71.  The
simulation was started from the static conductive state, beginning
with $R$ = 2,000.  After the steady-state was reached, the Rayleigh
number was raised step by step to higher values.  The Nusselt numbers
measured at the steady states are plotted in Fig.~\ref{fg:Nu} against
the Rayleigh number.  The simulation results of Clever and Busse
\cite{Clever74} are also plotted for comparison.  Agreement is found
at Rayleigh numbers less than 20,000.  At higher Rayleigh numbers, the
LBE simulation has a lower heat transport.  We have raised the
Rayleigh number to values as high as $10^6$ for the same resolution.
Unlike the thermal LBE model \cite{McNamara95}, the present model
remains numerically stable.

Shown in Figs.~\ref{fg:t} are the steady-state isotherms for some
typical Rayleigh numbers.  As the Rayleigh number is increased, the
temperature gradient near the boundary becomes sharper; the ascending
and descending fluid sheets become narrower, and the area in the
interior of the fluid with a reversed temperature gradient becomes
wider.

The steady-state solutions were obtained by raising the Rayleigh
number gradually after the convection roll has established at lower
Rayleigh number.  It was found however that if the simulation is
started from the static conductive state with $R$ = 50,000, the system
will evolve into an oscillatory state.  The dominant wavelength is
half of that in the steady states solutions shown in Figs.~\ref{fg:t},
and the ascending and descending fluid sheets swing back and forth
with a period of $0.174 L_z/v_c$.  The isotherms at the beginning, the
quarter, the half and three quarters of one oscillation period are
shown in Fig.~\ref{fg:osci}.  This oscillation does not occur in
simulations with $R\leq$ 30,000.

Three dimensional simulations were performed for the same Prandtl
number and Rayleigh numbers on a $128\times 128\times 32$ lattice.
Again, the computation was started from the static state with $R$ =
6,000.  Shown in Fig.~\ref{fg:Nu2} is the time history of the Nusselt
number as the Rayleigh number was raised step by step to the values
shown on the top of the graph.  Greyscale plots of temperature
distributions on the mid-plane at some typical times for different
Rayleigh numbers are shown in Figs.~\ref{fg:3d}.  The 2D convection
rolls have already exhibited some wavy instability at $R$ = 6,000.
However, the amplitude of the oscillation is so small that the
deformation of the convection rolls is difficult to be detected from
the static plots.  To reveal the details of the oscillation, the scale
has been enlarged and the time history of the Nusselt number replotted
in this section.  The slow decay of the amplitude of the oscillation
might be an indication that in agreement with other workers
\cite{Clever74,Mclaughlin82}, the Rayleigh number of 6,000 is very
close to the threshold at which the convection becomes time-dependent.
The evolution of the convection pattern became more and more
complicated and oscillations of more frequencies were involves as the
Rayleigh number was increased.  At $R$ = 30,000 and 50,000, although
the time history of the Nusselt number appeared to be quite chaotic,
the temperature field plotted in Figs.~\ref{fg:3d} still posses rather
regular structures and patterns.  Simulations at higher Rayleigh
number would require higher resolution.  A detailed investigation of
the transitions in RBC as the Rayleigh number is increased requires
large number of runs and is certainly beyond the scope of the present
work.

\section{Conclusions and Discussions}
\label{sec:4}

In this paper, we presented a method of simulating temperature
evolution in fluid systems using multiple component LBE model.  By
simulating the temperature field using an additional component, we are
able to avoid the numerical instability plaguing the thermal LBE
models.  The algorithm is simple, and the requirement on computational
resources is twice of that for a non-thermal LBE code.  As an example,
we studied the Rayleigh-B\'enard convection using this method.  The
results agree very well with theoretical predictions and experimental
observations both at near-critical and moderate Rayleigh numbers.

The density of the additional component satisfies a passive-scalar
equation.  In the simulation of the Boussinesq equations, the external
force is made to be a linear function of this passive scalar.
However, this passive-scalar can represent other properties of the
fluid satisfying more complicated equations.  More importantly, when
the equation of state is coupled with this passive scalar, the dynamic
process of phase transition can be simulated.  We defer the discussion
of the details to a future publication.

\section*{Acknowledgments}

The author is grateful to Drs.\ Francis Alexander, Gary Doolen, Nicos
Martys, David Montgomery, Joel Mozer, Rodney Worthing and Jeffrey
Yepez for helpful discussions.  The computation was performed using
the resources of the Advanced Computing Laboratory at Los Alamos
National Laboratory, Los Alamos, NM 87545.


\begin{table}
\caption{Critical Rayleigh number obtained by extrapolating growth
rate data at slightly supercritical Rayleigh numbers.}

\begin{tabular}{l|c|c|c|c|c|c}
Run \# & $L_y$ & $Pr$ & $\tau_1$ & $\tau_2$ & $R_c$ & Error\\ \hline
1 & 50 & 1 & 1 & 1 & 1707.11 & 0.04\% \\
2 & 30 & 1 & 1 & 1 & 1706.96 & 0.05\% \\
3 & 20 & 1 & 1 & 1 & 1706.87 & 0.05\% \\
4 & 10 & 1 & 1 & 1 & 1716.96 & 0.54\% \\ \hline
5 & 50 & 0.1 & 1 & 5.5 & 1715.75 & 0.47\% \\
6 & 50 & 100 & 1 & 0.505 & 1707.21 & 0.03\% \\
7 & 50 & 1 & 0.55 & 0.55 & 1713.84 & 0.36\% \\
8 & 50 & 1 & 1.5 & 1.5 & 1688.49 & 1.13\%
\end{tabular}
\label{tbl:rc}
\end{table}

\begin{figure}
\caption{Typical time-histories of the peak vertical velocity in 2D
simulation during the onset of the instability.  The Rayleigh numbers
are slightly above $R_c$.  Other parameters are: $Pr$ = 1, $\tau_1$ =
1, and the system size is 101 $\times$ 50.  The solid straight lines
are the drawn by least-square fitting, the slop of which gives the
growth rates of the instability.}
\label{fg:grow}
\end{figure}

\begin{figure}
\caption{Steady state isotherms and velocity field in a
two-dimensional simulation with the Rayleigh number of 1750.  The
resolution is 101 $\times$ 50.}
\label{fg:tnv}
\end{figure}

\begin{figure}
\caption{Growth rates of the instability are found to depend linearly
on the Rayleigh number near $R_c$.  The symbols are the results of
measurement from the time history of the peak vertical velocity, and
the straight lines are fitted through the data points.  The exact
critical Rayleigh number is obtained by extrapolating the data to the
zero growth rate.  Three sets of simulations with different values of
$\tau_1$ were performed to determine the effects of different $\tau_1$
on the accuracy.  The simulations were performed on a 2D 101 $\times$
50 lattice and is for $Pr$ = 1.}
\label{fg:rates}
\end{figure}

\begin{figure}
\caption{Time histories of the peak vertical velocity for different
system sizes.  It is to be seen that the simulation results converge
for system sizes $L_z >$ 20.  The peak velocity in the 3D simulation
has the same growth during the early development of the instability
and saturates at the same level.  The 3D effects peak around $t$ =
1320.}
\label{fg:conv}
\end{figure}

\begin{figure}
\caption{Greyscale plot of temperature distribution on the mid-plane
between the two walls at, (a) $t$ = 1047, (b) $t$ = 1320, (c) $t$ =
1524 and (d) $t$ = 2273.  The gray scales from the darkest to the
brightest are mapped to the temperature range 0.403 $< \theta <$
0.597.}
\label{fg:temp}
\end{figure}

\begin{figure}
\caption{The steady-state Nusselt number as function of the Rayleigh
number in two-D simulations.  The LBE results agree with that of
Clever and Busse for Rayleigh number less than 20,000.}
\label{fg:Nu}
\end{figure}

\begin{figure}
\caption{Two-D simulation.  Isotherms at steady states as the Rayleigh
number is raised gradually to (a) 10,000, (b) 20,000, (c) 30,000 and
(d) 50,000.}
\label{fg:t}
\end{figure}

\begin{figure}
\caption{Isotherms in two-D simulation.  The simulation was started
from the static conductive state with $R$ = 50,000.  The system
evolves into a oscillatory state.  The isotherms are taken at (a) the
beginning, (b) one quarter, (c) half, and (d) three quarter of one
oscillation period.}
\label{fg:osci}
\end{figure}

\begin{figure}
\caption{Time history of the Nusselt number in a 3D simulation
as the Rayleigh number is increased step by step at times indicated by
the vertical dashed lines.  For the first segment, the scale has been
adjusted to the range of 2.08 to 2.082 and the Nusselt number
replotted.}
\label{fg:Nu2}
\end{figure}

\begin{figure}
\caption{Greyscale plots of typical temperature distribution on the
mid-plane between the two walls for Rayleigh number (a) 6,000, (b)
8,000, (c) 10,000, (d) 20,000, (e) 30,000, and (f) 50,000.}
\label{fg:3d}
\end{figure}

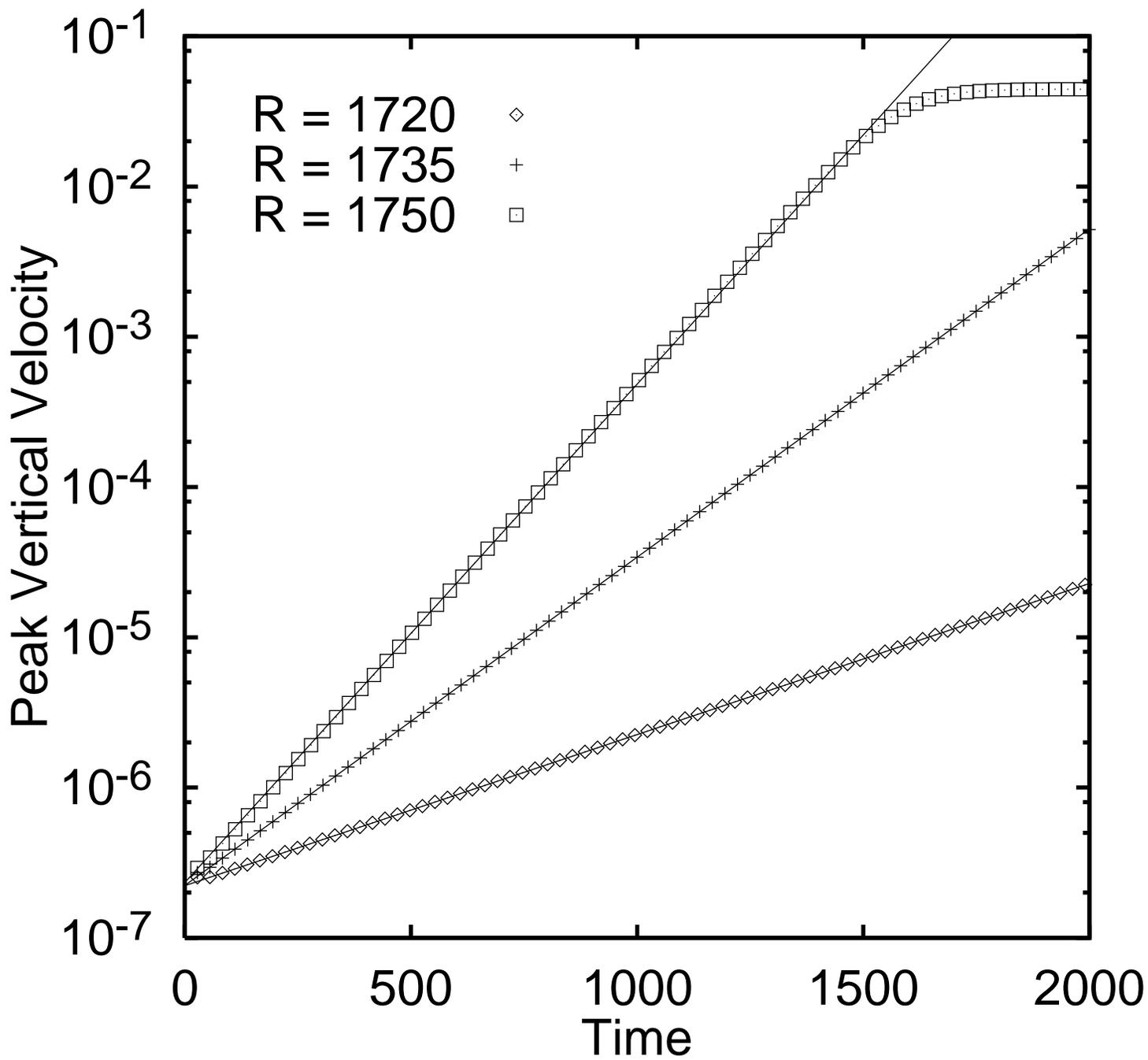
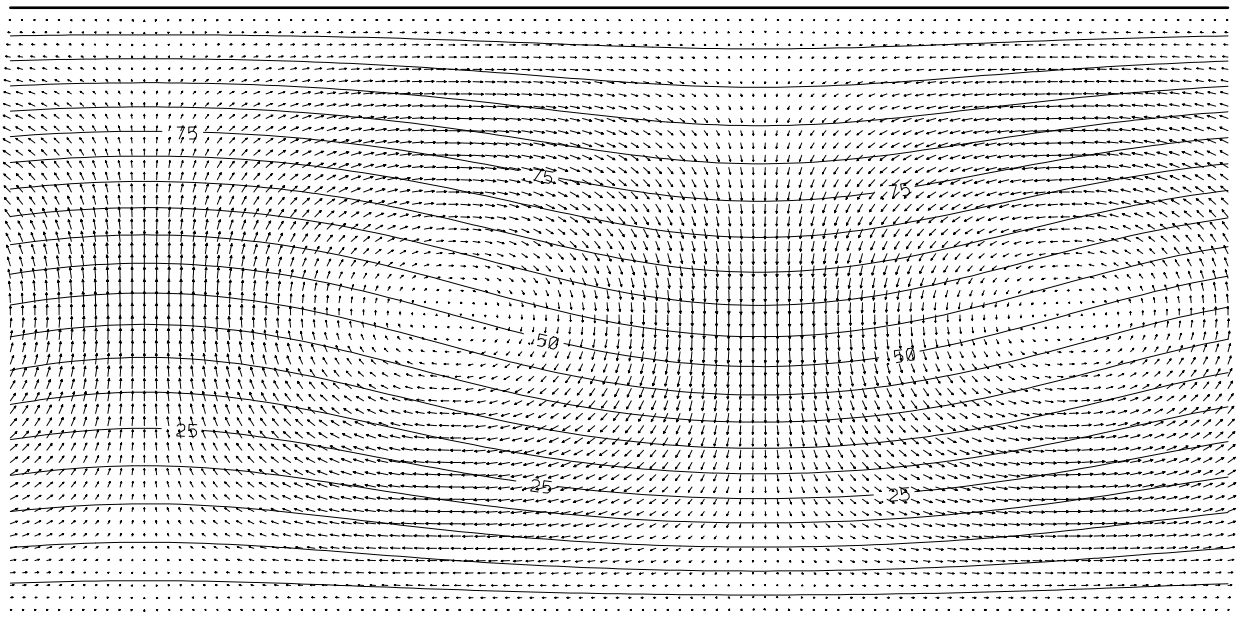
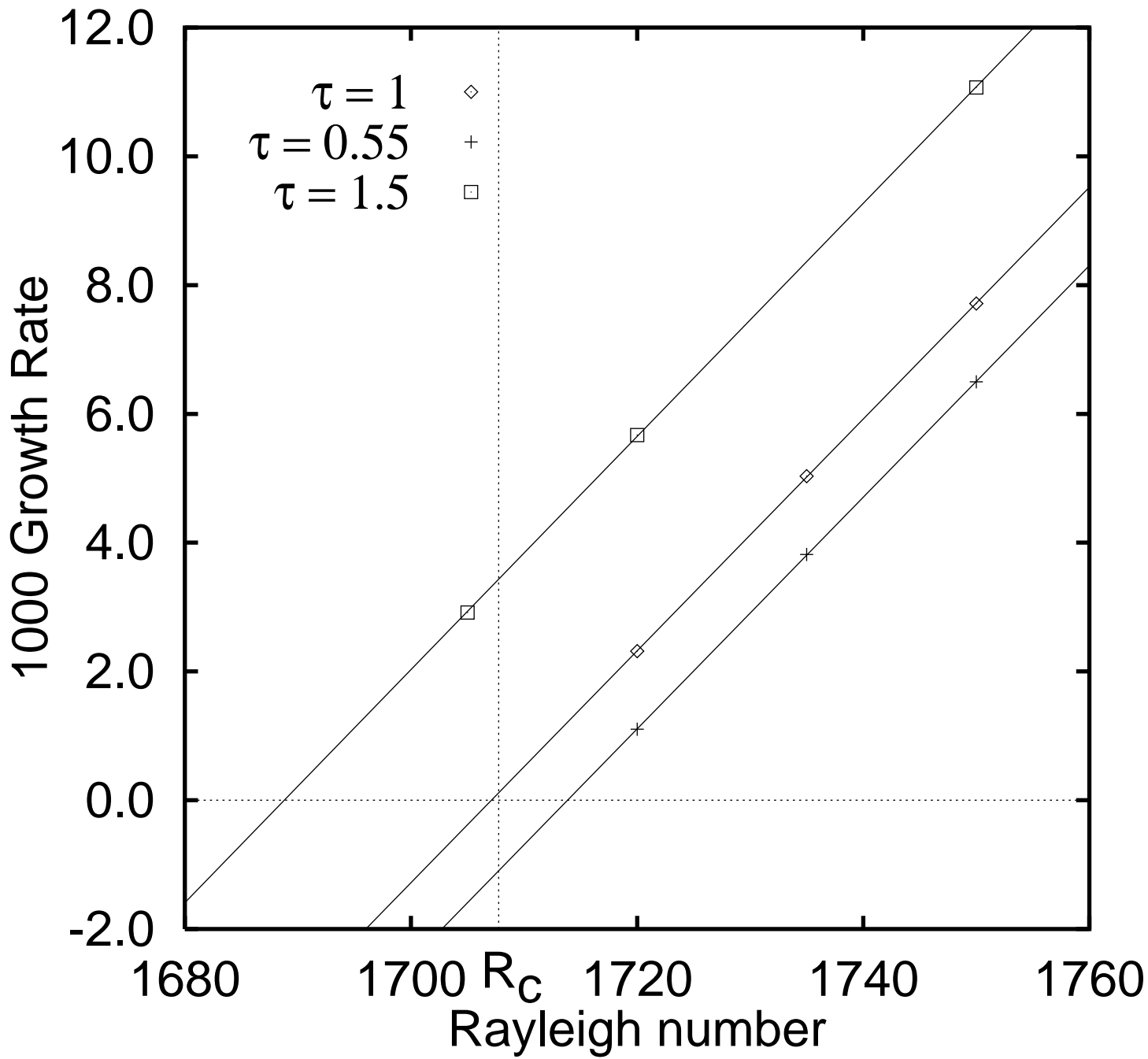
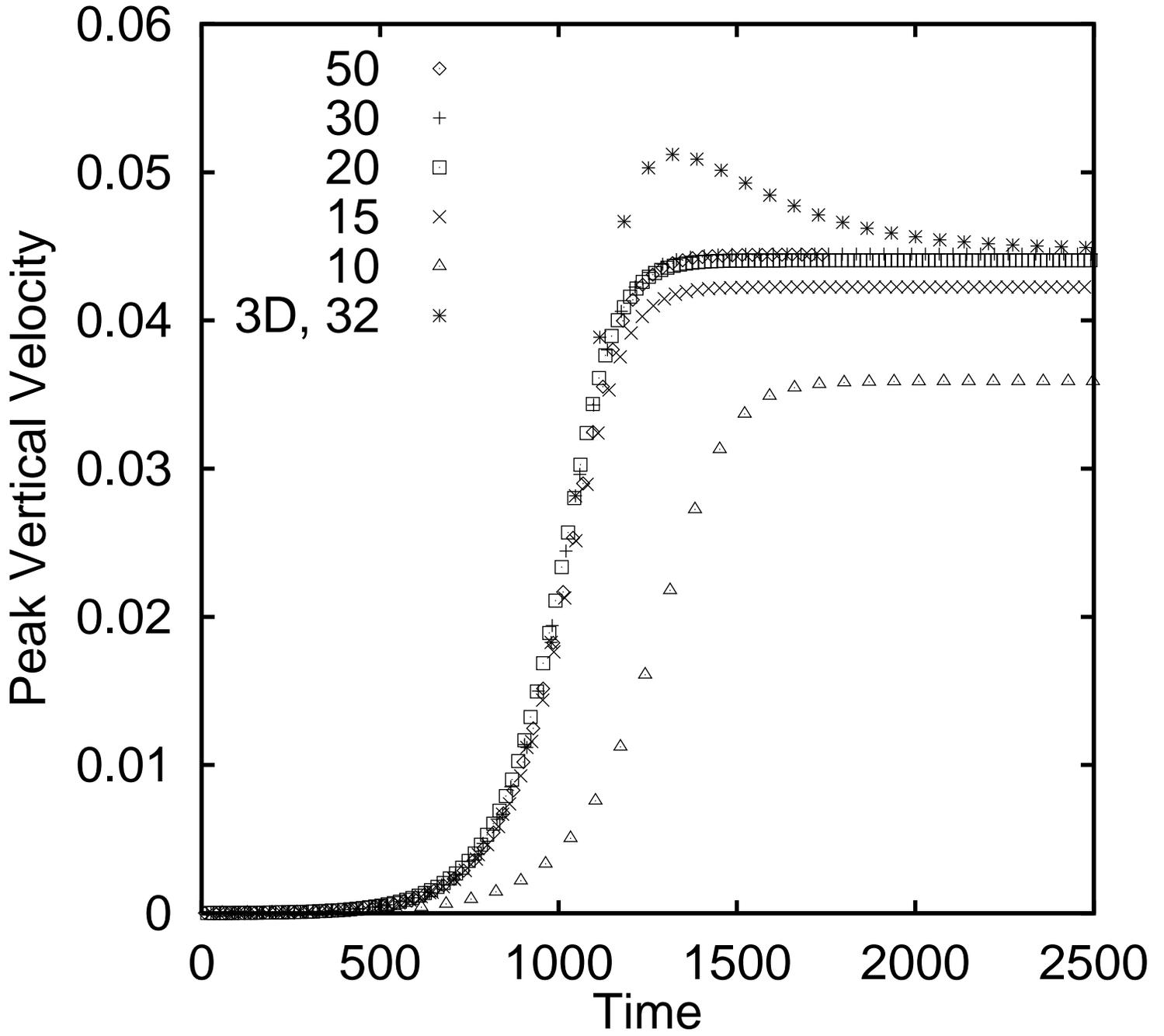
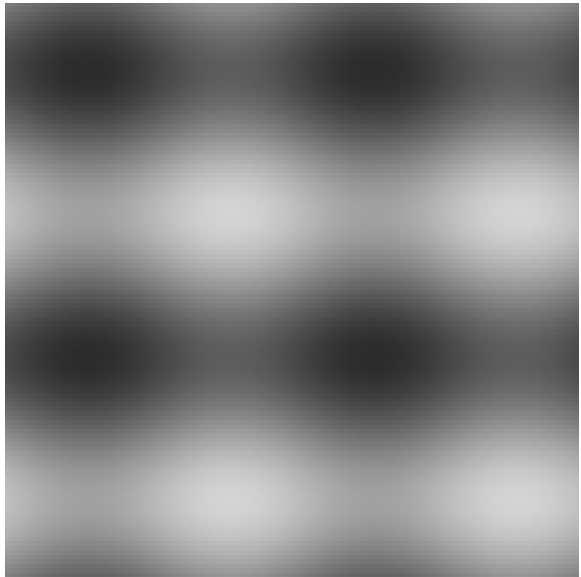
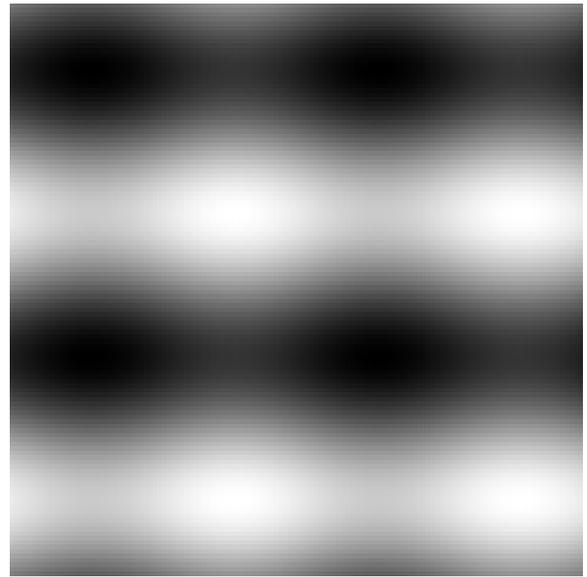
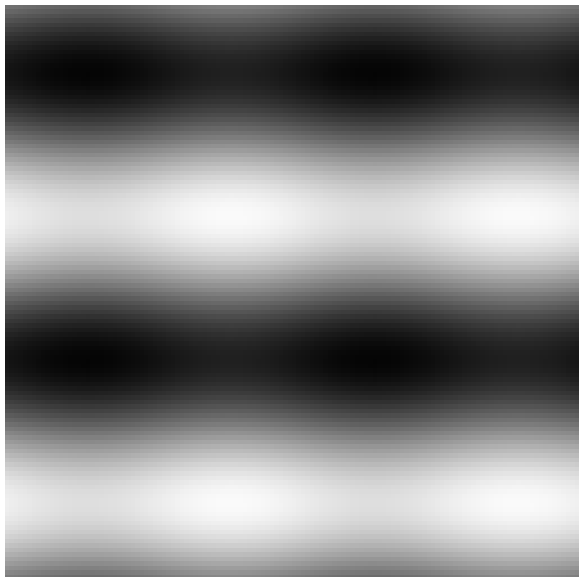
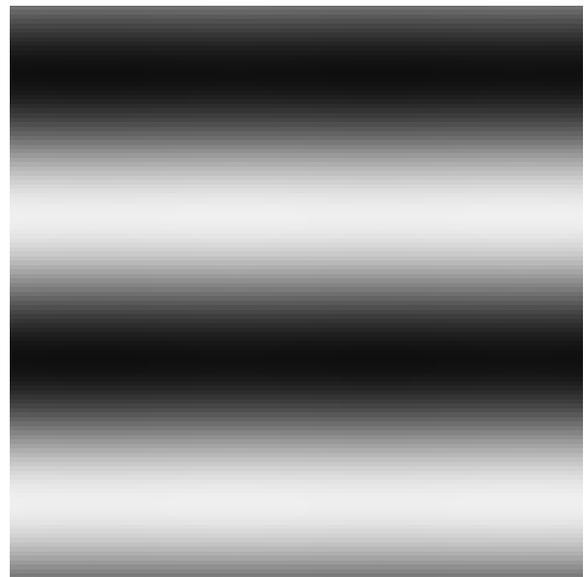
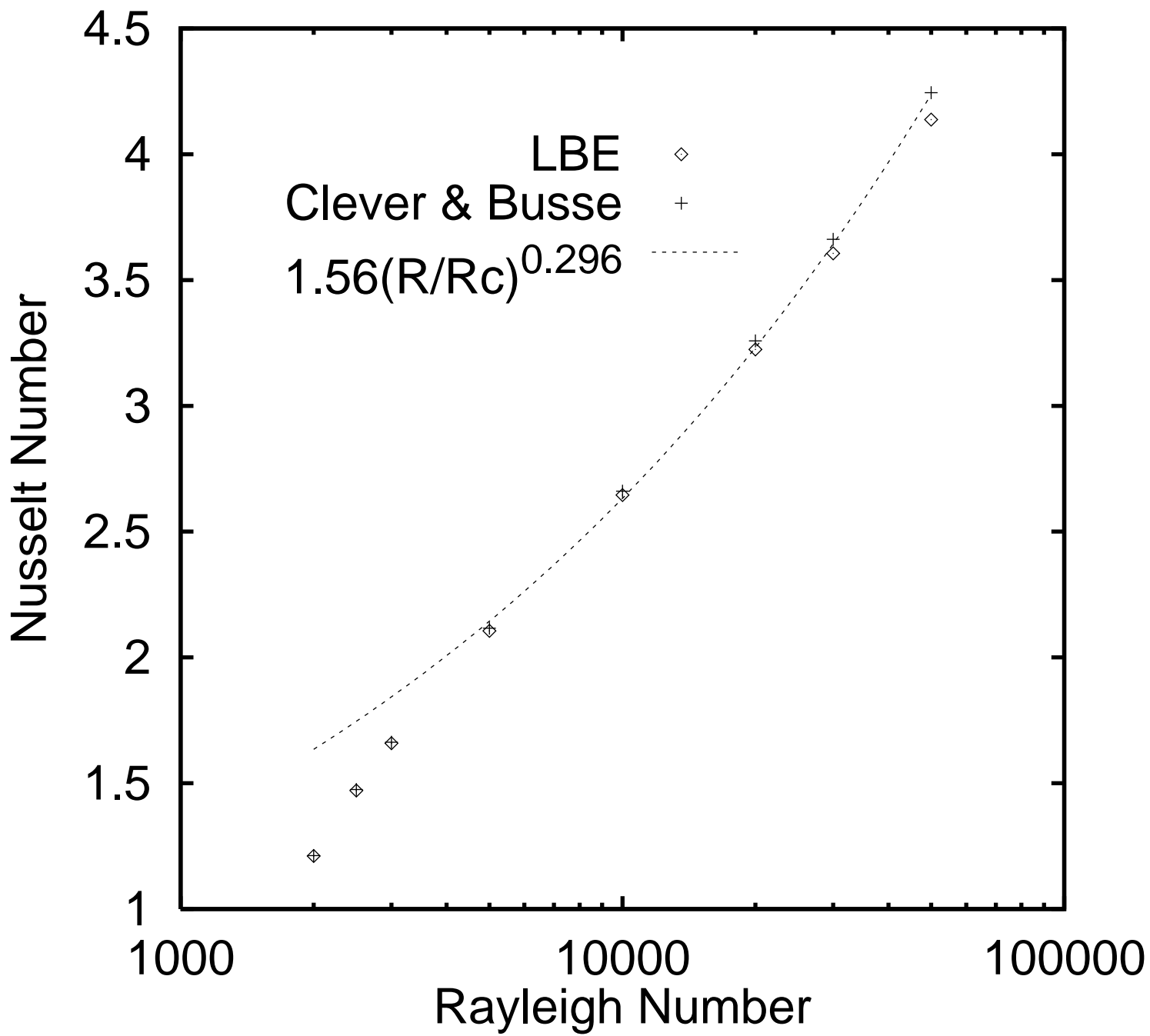
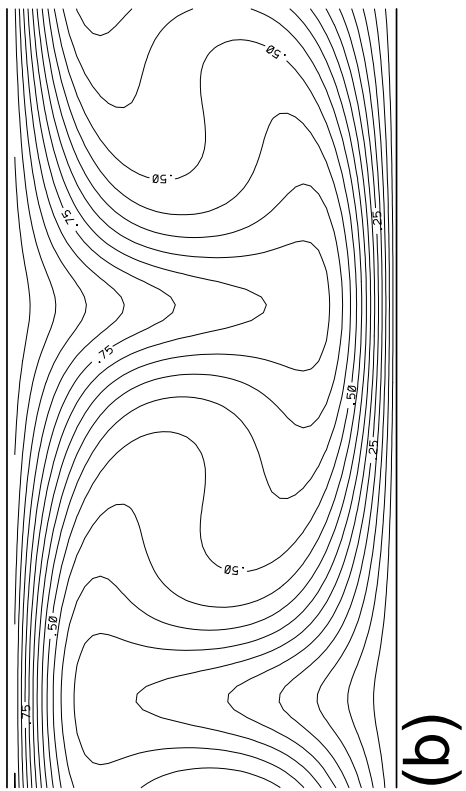
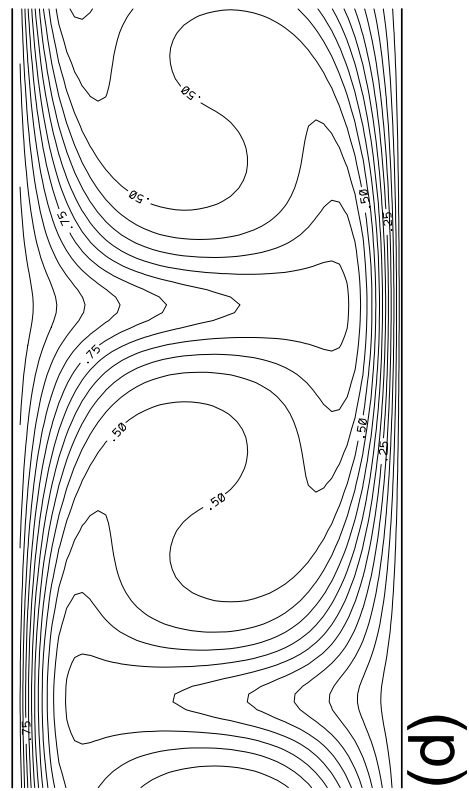
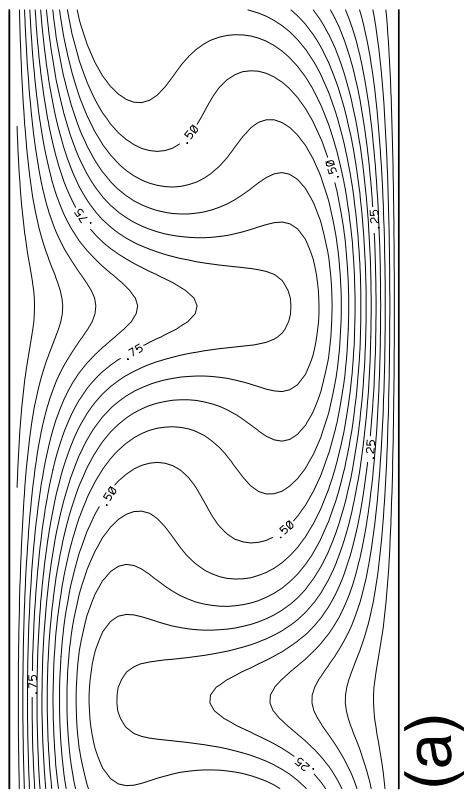
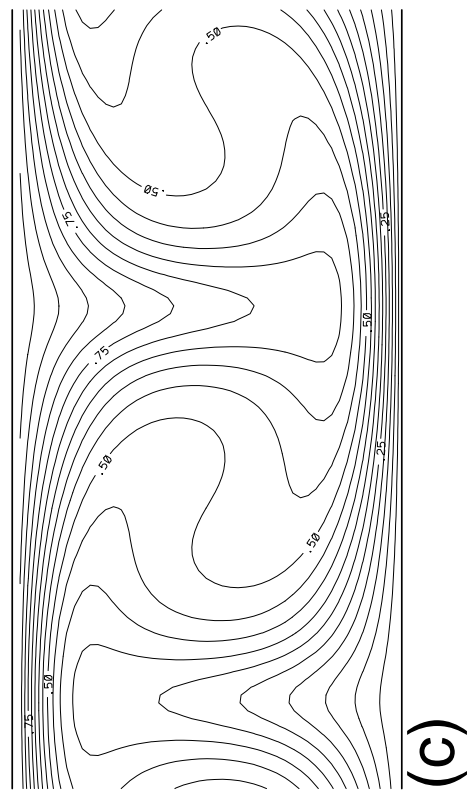
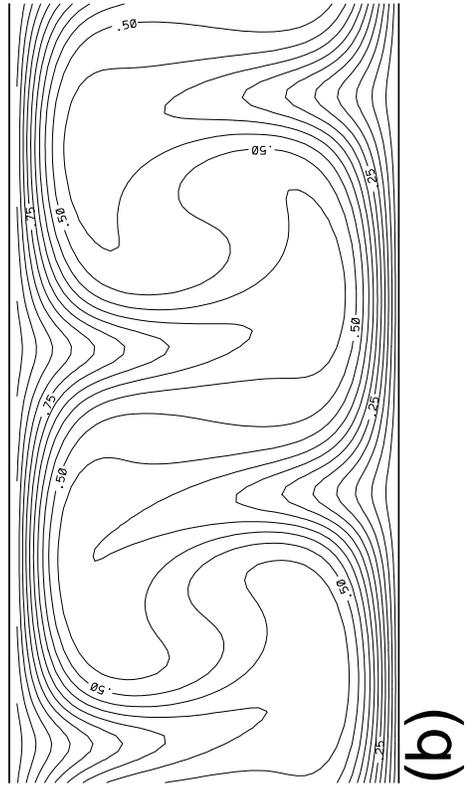
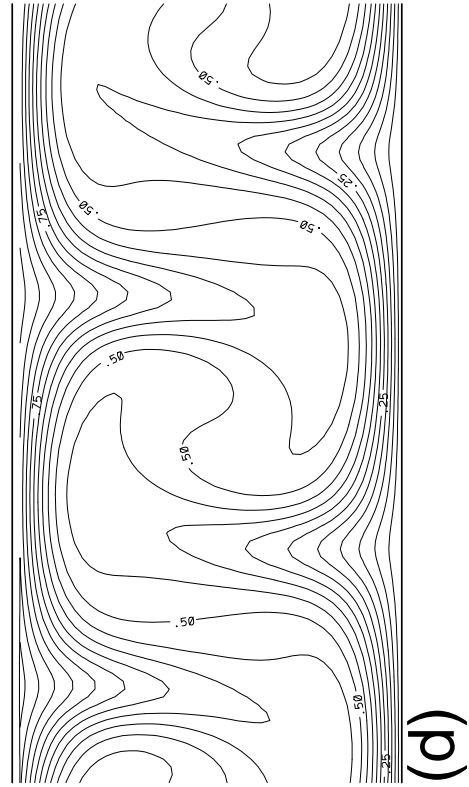
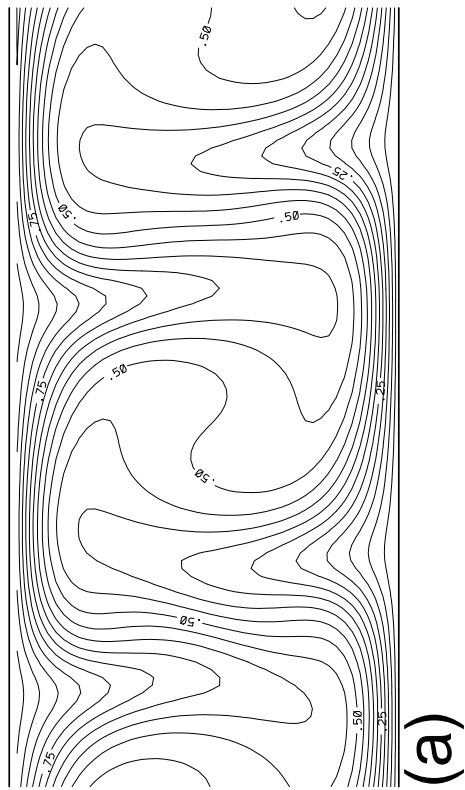
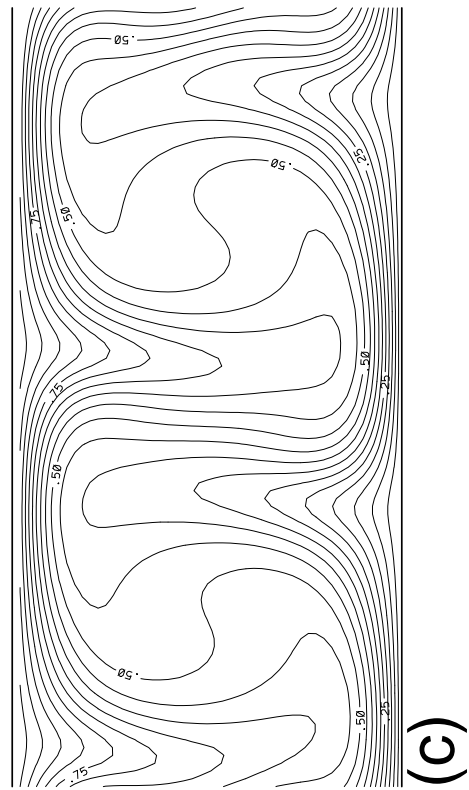
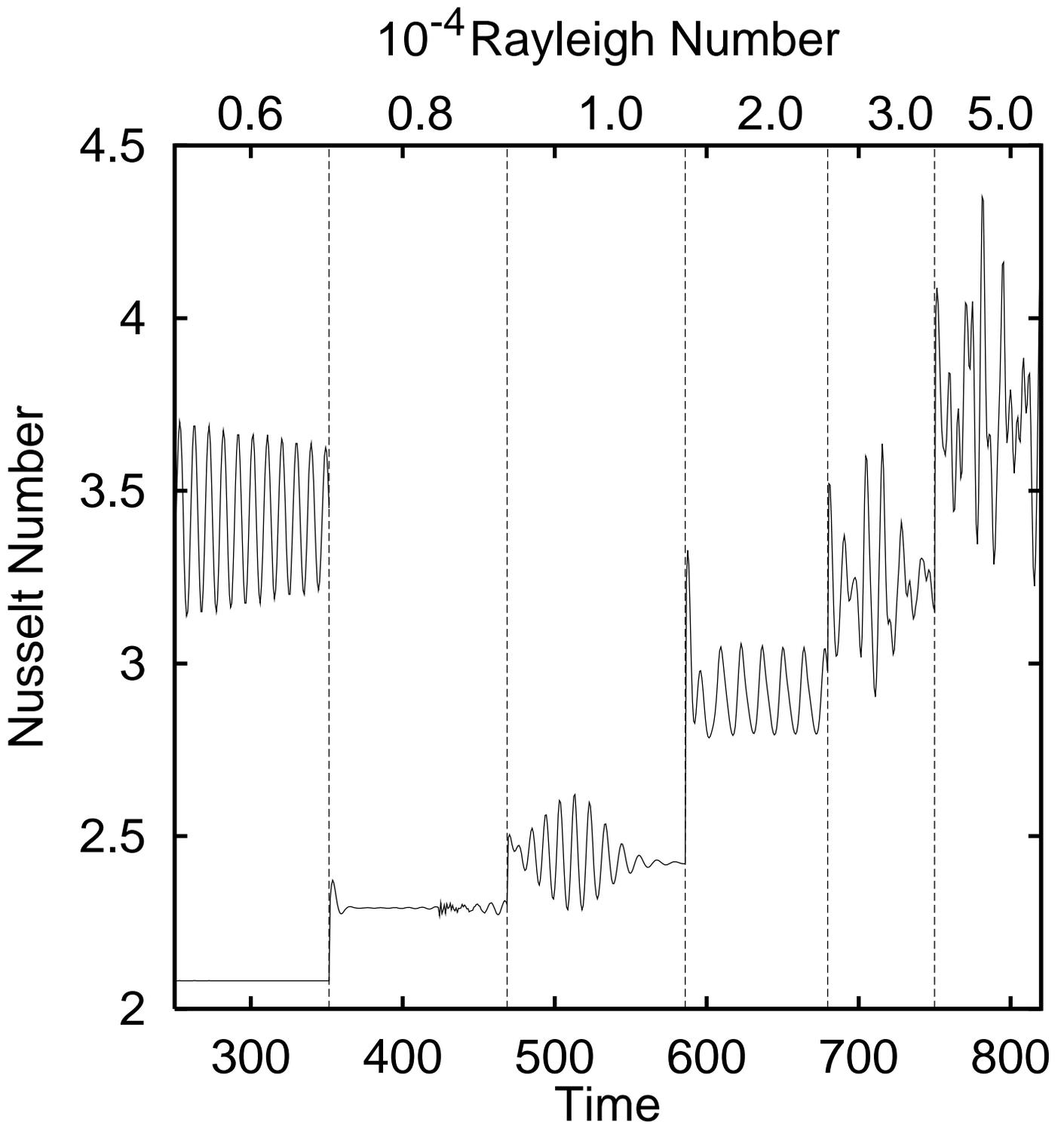
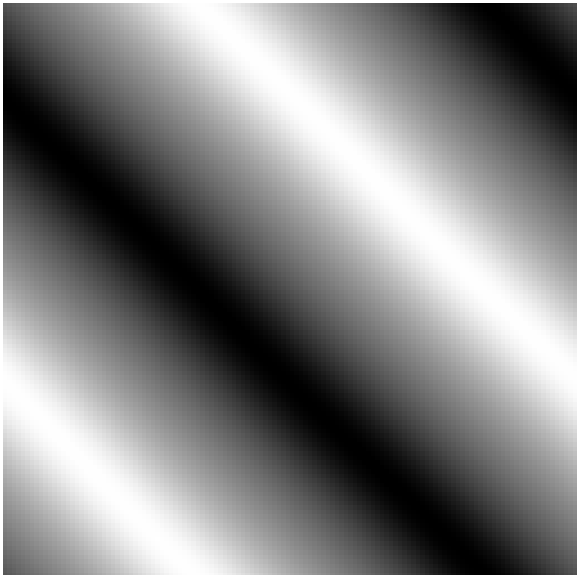
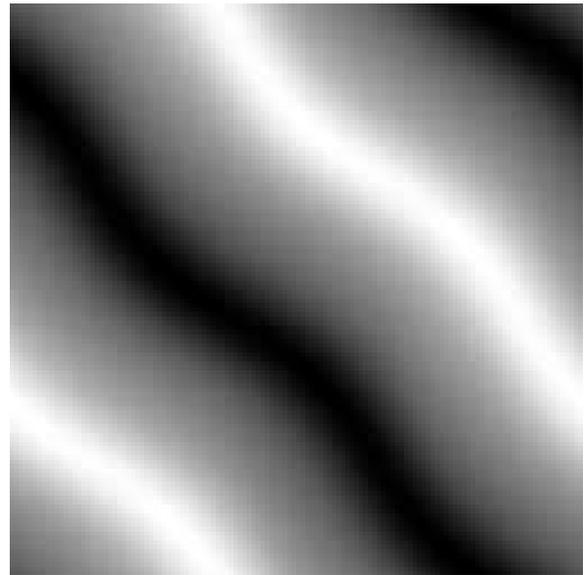
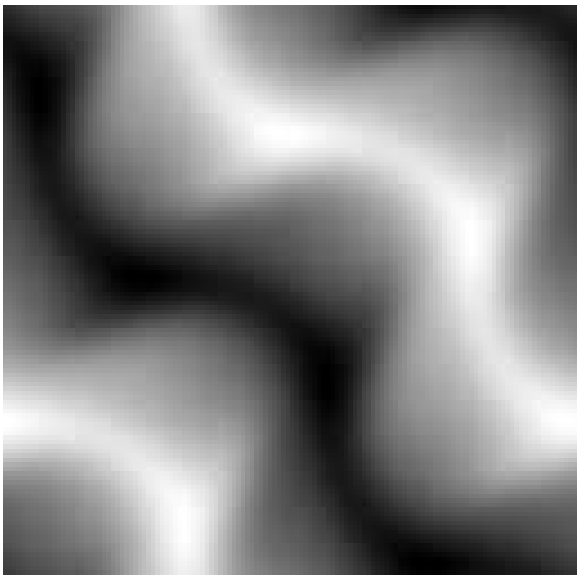
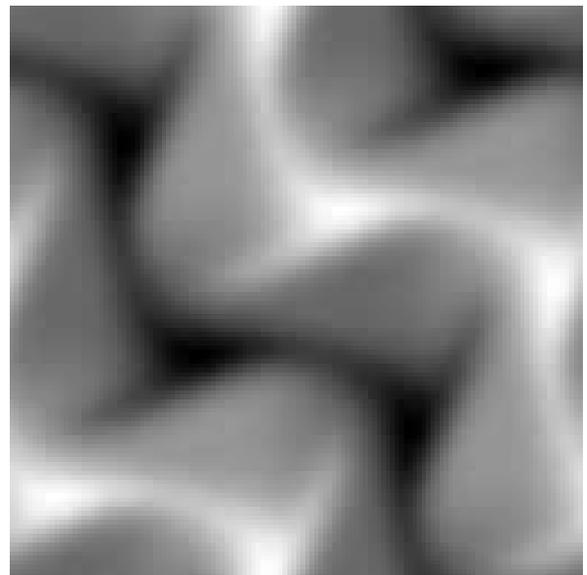
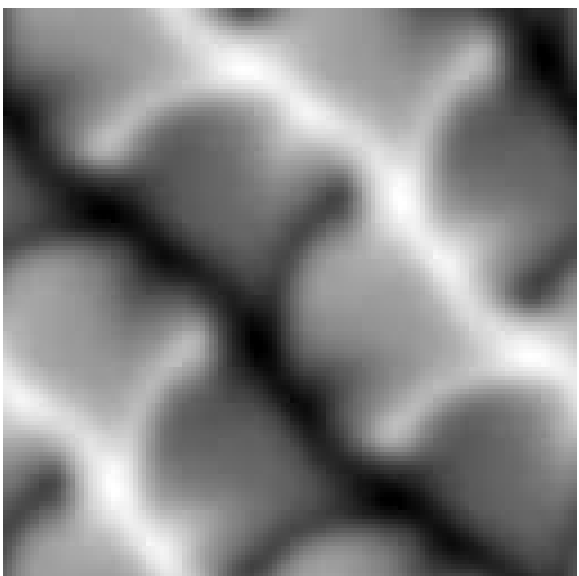
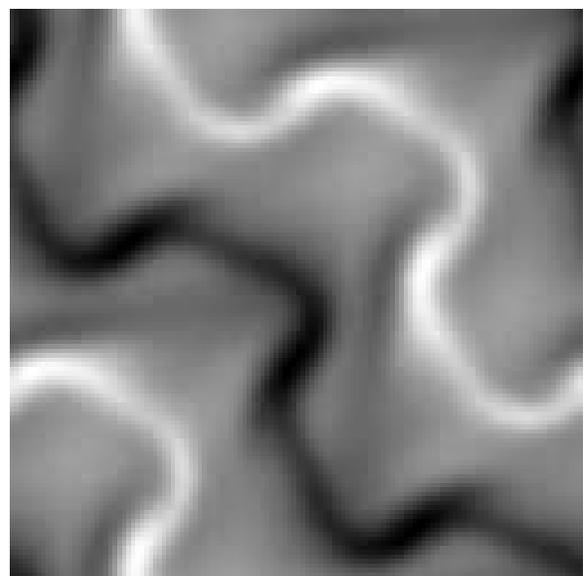
\end{document}